\documentclass{svmult}
\usepackage{amssymb,amsfonts,latexsym,stmaryrd}
\usepackage{QED}
\usepackage[PostScript=dvips]{diagrams}
\usepackage{euscript}
\usepackage{pstricks,pst-node,pst-tree}
\usepackage{epsfig}
\usepackage{url}

\newcommand{\beqa}{\begin{eqnarray*}}
\newcommand{\eeqa}{\end{eqnarray*}\par\noindent}

\renewcommand{\emph}[1]{\textbf{#1}}

\newcommand{\true}{\mathbf{tt}}
\newcommand{\false}{\mathbf{ff}}

\newcommand{\rarr}{\rightarrow}

\newcommand{\ie}{\textit{i.e.}~}

\newcommand{\pfn}{\rightharpoonup}

\newarrow{Eq}=====
\newarrow{mon}>--->
\newarrow{rel}--+->
\newarrow{inc}C--->
\newcommand{\graph}{\mathsf{graph}}

\newcommand{\Bool}{\mathbb{B}}

\newcommand{\liff}{\leftrightarrow}
\newcommand{\Nat}{\mathbb{N}}
\newcommand{\IMPLIES}{\; \Rightarrow \;}
\newcommand{\sqle}{\sqsubseteq}
\newcommand{\lfp}[1]{\mathsf{lfp}(#1)}
\newcommand{\olub}{\bigsqcup_{n \in \omega}}

\newcommand{\por}{\mathsf{por}}
\newcommand{\lsor}{\mathsf{lsor}}
\newcommand{\rsor}{\mathsf{rsor}}
\newcommand{\Cond}[3]{\mathsf{if} \; #1 \; \mathsf{then} \; #2 \; \mathsf{else} \; #3}
\newcommand{\Opp}{{\color{blue} O}}
\newcommand{\Pla}{{\color{red} P}}

\newcommand{\GG}{\mathcal{G}}

\renewcommand{\color}[1]{}

\begin{document}

\title{Intensionality, Definability and Computation}
\author{Samson Abramsky}
\institute{Department of Computer Science, University of Oxford \at Wolfson Building Building, parks Road, Oxford OX1 3QD, U.K.
\email{samson.abramsky@cs.ox.ac.uk}
}
\date{}

\maketitle

\begin{abstract}
\_ We look at intensionality from the perspective of computation.
In particular, we review how game semantics has been used to characterize the sequential functional processes, leading to powerful and flexible methods for constructing fully abstract models of programming languages, with applications in program analysis and verification.
In a broader context, we can regard game semantics as a first step towards developing a 
positive theory of intensional structures with a robust mathematical structure, and finding 
the right notions of invariance for these structures.
\end{abstract}


\section{Introduction}

Our aim in this paper is to give a conceptual discussion of some issues concerning intensionality, definability and computation.
Intensionality remains an elusive concept in logical theory, but actually becomes much more tangible and, indeed, inescapable in the context of computation. We will focus on a particular thread of ideas, leading to recent and ongoing work in game semantics. Technical details will be kept to a minimum, while ample references will be provided to the literature.

It is a pleasure and a privilege to contribute to a volume in honour of Johan van Benthem. Johan has been a friend and an inspiring and supportive colleague for well over a decade.
He has encouraged me to broaden my intellectual horizons and to address wider conceptual issues.
Although our technical languages and backgrounds are rather different, we have found a great deal of common ground in pursuing a broad and inclusive vision of logical dynamics. I look forward to continuing our interactions and discussions over the coming years.

\subsection{Computability vs. Computer Science}
Computability theory \cite{rogers1987theory,cutland1980computability} is concerned with the computability (or, most often, degree of \emph{non}-computability) of \emph{extensional objects}: numbers, sets, functions etc. These objects are inherited from mathematics and logic.
To define when such objects are computable requires some additional structure: we say e.g.~that a function is computable if there exists an \emph{algorithmic process} for computing it. Hence the notion of computability  relies on a characterization of algorithmic processes. This was, famously, what Turing achieved in his compelling analysis \cite{turing1938computable}.

Computer Science asks a broader question:
\begin{center}
\fbox{What is a process?}
\end{center}

By contrast with the well-established extensional notions which provide the reference points for computability, there was no established mathematical theory of what processes are predating computer science.

\subsection{Why processes matter in Computer Science}
Let us pause to ask why Computer Science asks this broader question about the nature of informatic processes.
The purpose of much of the software we routinely run is not to compute a function, but to \emph{exhibit some behaviour}.
Think of communication protocols, operating systems, browsers, iTunes, Facebook, Twitter, \ldots .
The purpose of these systems is not adequately described as the computation of some function.

Thus we are led ineluctably to questions such as:
\begin{center}
What is a process? When are two processes equivalent?
\end{center}

The situation is very different to that which we find in computability theory, where we have
\begin{itemize}
\item A confluence of notions, whereby many different attempts to characterize the notion of algorithmic process have converged to yield the same class of computable functions \cite{rogers1987theory,cutland1980computability}.
\item A definitive calculus of functions: the $\lambda$-calculus \cite{church1941calculi,barendregt1984lambda}.
\end{itemize}

There has been active research on concurrency theory and processes for the past five decades in Computer Science \cite{petri1966communication,petri1962fundamentals,milner1989communication,milner1999communicating,hoare1978communicating}.
Many important concepts and results have emerged. However, one cannot help noticing that:
\begin{itemize}
\item Hundreds of different process calculi, equivalences, logics have been proposed.
\item No $\lambda$-calculus for concurrency has emerged.
\item There is no plausible Church-Turing thesis for processes.
\end{itemize}
This has been referred to as the `next 700' syndrome, after Peter Landin's paper (from 1966!) on `The Next 700 Programming Languages' \cite{landin1966next}; for a discussion of this syndrome, see \cite{abramsky2006fundamental}.
Finding an adequate characterization of informatic processes in general
can plausibly be considered to be a much harder problem than that of characterizing the notion of computable set or function. Despite the great achievements of Petri, Milner, Hoare,  et al., we still await our modern-day Turing to provide a definitive analysis of this notion.

\subsection{Prospectus}
Our aim in this paper is to tell one limited but encouraging success story: the characterization of  \emph{sequential functional processes} using \emph{game semantics}, solving in best possible terms the `full abstraction problem' for PCF \cite{milner1977fully,plotkin1977lcf}.
This has led on to many further developments, notably :
\begin{itemize}
\item Full abstraction and full completeness results for a wide range of programming languages, type theories and logics \cite{abramsky1994games,abramsky2000full,hyland2000full,abramsky1997linearity,abramsky1999game,abramsky1998fully,abramsky1999concurrent,abramsky1998call,honda1997game,laird2001fully,laird2005game}.
\item A basis for compositional program verification \cite{abramsky2004applying,ghica2009applications}.
\end{itemize}
There is much ongoing work \cite{murawski2009full,murawski2011game,
murawski2011algorithmic}, and this continues to be a flourishing field.\footnote{See e.g. \url{https://sites.google.com/site/galopws/} for a workshop series devoted to this topic.}

In a broader context, we can regard game semantics as a first step towards developing a positive theory of intensional structures with a robust mathematical structure, and finding the right notions of invariance.

\section{Intensionality vs. Extensionality}

The notions of intensionality and extensionality carry symmetric-sounding names, but this apparent symmetry is misleading. Extensionality is enshrined in mathematically  precise axioms with a clear conceptual meaning. Intensionality, by contrast, remains elusive. It is a ``loose baggy monster''\footnote{Cf. Henry James on the Russian masters.} into which all manner of notions may be stuffed, and a compelling and coherent general framework for intensional concepts is still to emerge.

Let us recall some basic forms of extensionality.
For sets we have:
\[ x = y \;\; \liff \;\; \forall z. \, z \in x \liff z \in y . \]
This says that a set is completely characterized by its \emph{members}.

For functions we have:
\[ f = g \;\; \liff \;\; \forall x. \, f(x) = g(x) . \]
This says that a function is completely characterized by the \emph{input-output correspondence} (\ie the set of 
input-output pairs) which it defines.

The common idea underlying these principles is that mathematical entities can be characterized in \emph{operational} terms: mathematical objects should be completely determined by their \emph{behaviour} under the operations which can be applied to them. In the case of sets, this operation is testing elements for membership of the set, while in the case of functions, it is applying functions to their arguments.

The basic question we are faced in seeking to make room for intensional notions  is this:
is intensionality just a \emph{failure} to satisfy such properties?
Or is there some \emph{positive story} to tell?

We shall focus on the case of functions. Here we can say that the modern, extensional view of functions as completely determined by their graphs of input-output correspondences over-rode an older, intensional notion, of a function being given by its \emph{rule}.
That older notion was never adequately formalized. Much of modern logic, in its concern with issues of definability, can be seen as providing tools for capturing the old intuitions in a more adequate fashion.

\subsection{Intrinsic vs. Extrinsic Properties of Functions}
We shall now draw a distinction which will be useful in our discussion.
We say that a  property of functions
\[ f : A \rarr B \]
is \emph{intrinsic} if it can be defined  purely in terms of $f$ (as a set of input-output pairs) and any structure pertaining to $A$ and $B$.

This is, of course, not very precise.
More formally, we could say: if it can be defined using only bounded quantification over the structures $A$ and $B$.
But we shall rest content with the informal rendition here. We believe that the distinction will be quite tangible to readers with some mathematical experience.

\subsection{Examples}

\begin{itemize}
\item $A$ and $B$ are groups. A function $f : A \rarr B$ is a \emph{group homomorphism} if
\[ f(xy) = f(x)f(y), \qquad f(e) = e . \]
In general, homomorphisms of algebraic structures are clearly intrinsic in the sense we intend.
\item $A$ and $B$ are topological spaces.  A function $f : A \rarr B$ is \emph{continuous} if $f^{-1}(U)$ is open in the topology on $A$ for every open subset $U$ of $B$. Here the topology is viewed as part of the structure.
\end{itemize}

\subsection{A Non-example: Computability}

Computability is \emph{not} an intrinsic property of partial functions
\[ f : \Nat \pfn \Nat \]
in this sense.
In order to define whether a function is computable, we need to refer to something \emph{external}; a \emph{process} by which $f$ is computed.
A function is computable if there is some \emph{algorithmic process} which computes $f$.
But what is an algorithmic process?

Here of course we can appeal to
Turing's analysis \cite{turing1938computable}, and to subsequent, axiomatic studies by Gandy et al. \cite{gandy1980church,sieg2002calculations}.
But note that, not only do we have to appeal to some external notion of machine or algorithmic process, but
there is \emph{no single canonical form} of external structure witnessing computability.
Rather, we have a \emph{confluence}: all `reasonable' notions lead to the same class of computable functions. But this confluence still leaves us with an extrinsic definition, and moreover one in which the external witness has no canonical form.

More concretely, suppose we are given some function
\[ f : \{ 0, 1 \}^* \rTo \{ 0, 1 \} \]
\ie a predicate on binary strings.
To say that whether such a function is computable is an extrinsic property of $f$ simply means that we cannot say if $f$ is computable just by looking at its input-output graph and properties relating to the structure of $\{ 0, 1 \}^*$ and $\{ 0, 1 \}$.
Clearly, we need something more, typically either: 
\begin{itemize}
\item A suitable notion of \emph{machine}, or

\item An inductive definition given by some `function algebra' or logical theory.
\end{itemize}

\subsection{A comparison point: regular languages}

One might reasonably ask what it would even mean to have an intrinsic (or, at least, \emph{more} intrinsic) way of defining computability. For this purpose, it is useful to consider a much simpler notion which pertains to (sub-)computability in a non-trivial fashion, and which \emph{does} admit an intrinsic definition in our sense.

Such an example is provided by regular languages, \ie those accepted by finite-state automata \cite{hopcroft1979introduction}.

Let $A$ be some finite alphabet. We write $A^*$ for the set of finite words or strings over this alphabet.
Given a language $L \subseteq A^*$, we define
\[ s \equiv_L t \;\; \liff \;\; \forall v \in A^* . \, sv \in L \liff tv \in L . \]
Clearly, $\equiv_L$ is an equivalence relation on $A^*$.

\begin{theorem}[Myhill-Nerode \cite{myhill1957finite,nerode1958linear}]
$L$ is regular if and only if $\equiv_L$ is of finite index (\ie has finitely many distinct equivalence classes).
\end{theorem}

Of course, computability is a much richer notion than regularity, and one may well suppose that for metamathematical reasons, no intrinsic or quasi-intrinsic definition of computability can be achieved --- although we are not aware of any specific formal result which implies this.

Still, the question seems worth asking: we suspect that a better understanding of this issue may be important in addressing some fundamental questions in computability and complexity. We shall return briefly to this point in the concluding section.

\section{From functions to functionals}

The issues become clearer if we include \emph{functionals} (functions which take functions as arguments) in our discussion.
This leads to the following hierarchy of types:

\begin{center}
\begin{tabular}{cl}
Type $0$: & $\Nat$\\
Type 1:  & $\Nat \pfn \Nat$\\
Type 2:  & $[\Nat \pfn \Nat] \pfn \Nat$ \\
$\vdots$ & \\
\end{tabular}
\end{center}
In general, a type $n+1$ functional takes type $n$ functionals as arguments.
Of course, more general types can also be considered.

Functionals are not so unfamiliar: e.g. the quantifiers!
\[ \forall, \exists : [\Nat \rarr \Bool] \rarr \Bool \]
Here $\Bool$ is the set  of truth-values; the quantifiers over the natural numbers are seen as functionals taking natural number predicates to truth-values.

While it might seem that bringing functionals into the picture will merely complicate matters, in fact when we consider higher-order functions, some intrinsic structure emerges naturally, which is lacking when we only look at first-order functions over discrete data.

When we compute with a function (or procedure) parameter $P$, we can immediately distinguish two paradigms.
\begin{itemize}
\item Extensional paradigm: the only way we can interact with $P$ is to call it with some arguments, and use the results:
\[ \mathsf{let} \; m = P(n) \; \mathsf{in} \; \ldots \]
In any finite computation, we can only make finitely many such calls, and hence `observe' a finite subset of the graph of the function defined by $P$:
\[ \{ m_1 = P(n_1), m_2 = P(n_2), \ldots , m_k = P(n_k) \} \]
\item Intensional paradigm: we have access to the \emph{code} of $P$.
So we can compute such things as the code itself (as a text string), or how many symbols appear in it, etc.
Examples: interpreters, program analyzers, and other \emph{programs which manipulate programs}.
Usually, though, in computer science we keep programs as data (subject to manipulation) distinct from programs as code (performing manipulations); it is generally seen as bad practice to mingle these two modes. By contrast, computability theory does use codes of programs viewed as data in a pervasive and essential fashion. We shall return to this contrast when we discuss fixpoint theorems.
\end{itemize}

\subsection{Intrinsic structure of computable functionals}

When we pass to functionals, some intrinsic structure begins to emerge.
Firstly, there is a natural ordering on partial functions:
\[ f \sqsubseteq g \;\; \liff \;\; \graph(f) \subseteq \graph(g) . \]
Here $\graph(f)$ is the set of input-output correspondences which --- on the extensional view --- uniquely characterize $f$. If $P$ is a program code, we can define $\graph(P)$ to be the set of input-output correspondences defined by the (partial) function computed by $P$.

We can view a function which takes codes of programs and returns numbers as an (\textit{a priori} intensional) functional. We say that such a function $F$ is extensional if  $F(P) = F(Q)$ whenever $\graph(P) = \graph(Q)$, and hence defines 
a functional $\hat{F} : [\Nat \pfn \Nat] \pfn \Nat$.

We have the following classical result:
\begin{theorem}[Myhill-Sheperdson \cite{myhill1955effective}]
An extensional computable functional $\hat{F}$ satisfies the following properties:
\begin{description}
\item[Monotonicity] : $f \sqsubseteq g \IMPLIES \hat{F}(f) \sqsubseteq \hat{F}(g)$.
\item[Continuity] : For any increasing sequence of partial functions
\[ f_0 \sqsubseteq f_1 \sqsubseteq f_2 \sqsubseteq \cdots \]
we have
\[ \hat{F}(\bigsqcup_i f_i) \; = \; \bigsqcup_i \hat{F}(f_i) . \]
\end{description}
Moreover, there is a sort of converse (which still needs to appeal to the usual notion of computability for functions on the natural numbers).
\end{theorem}

\subsection{Generalization to Domains}
These ideas were elaborated into a beautiful mathematical theory of computation by Dana Scott and others.
Domains are partial orders with least elements and least upper bounds of increasing sequences.
The corresponding functions are the monotonic and continuous ones.

A function $f : D \rightarrow E$ is \emph{monotonic} if, for all $x, y \in D$:
\[ x \sqle y \;\; \Longrightarrow \;\; f(x) \sqle f(y) . \]
It is \emph{continuous} if it is monotonic, and for all $\omega$-chains $(x_{n})_{n \in \omega}$ in $D$:
\[ f (\bigsqcup_{n \in \omega} x_{n}) = \bigsqcup_{n \in \omega} f(x_{n}) . \]
Continuity serves as an `intrinsic approximation' to computability: an intrinsic property which is a necessary condition for computability. Indeed, one speaks of `Scott's thesis', that computable functions are continuous \cite{winskel1989introduction}. Of course, this condition is not sufficient, and hence does not offer a complete analysis of computability.

\paragraph{Examples}
We consider functions $f : \{ 0, 1 \}^{\infty} \rightarrow \Bool_{\bot}$.
Here $\{ 0, 1 \}^{\infty}$ is the domain of finite and infinite binary sequences (or `streams'), ordered by prefix; while $\Bool_{\bot} = \{ \true, \false, \bot \}$ is the `flat' domain of booleans with $\true \sqsupseteq \bot \sqsubseteq \false$, representing computations which either fail to halt, or return a boolean value.

We consider the following definitions for such functions:
\begin{enumerate}
\item $f(x) = \true$ if $x$ contains a 1, $f(x) = \bot$ otherwise.
\item  $f(x) = \true$ if $x$ contains a 1, $f(0^{\infty}) = \false$, $f(x) = \bot$ otherwise.
\item  $f(x) = \true$ if $x$ contains a 1, $f(x) = \false$ otherwise.
\end{enumerate}
Of these: (1) is continuous, (2) is monotonic but not continuous, and (3) is not monotonic.

The conceptual basis for monotonicity is that the information in Domain Theory is \emph{positive}; negative information is not regarded as stable observable information. That is, if we are at some information state $s$, then for all we know, $s$ may still increase to $t$, where $s \sqle t$. This means that if we decide to produce information $f(s)$ at $s$, then we must produce all this information, and possibly more, at $t$, yielding $f(s) \sqle f(t)$. Thus we can only make decisions at a given  information state which are stable under every possible information increase from that state. This idea is very much akin to the use of partial orders in Kripke semantics for Intuitionistic Logic \cite{kripke1965semantical}, in particular in connection with the interpretation of negation in that semantics.

The continuity condition, on the other hand, reflects the fact that a computational process will only have access to a finite amount of information at each finite stage of the computation. If we are provided with an infinite input, then any information we produce as output at any finite stage can only depend on some finite observation we have made of the input. This is reflected in one of the inequations corresponding to continuity:
\[ f (\bigsqcup_{n \in \omega} x_{n}) \sqsubseteq \bigsqcup_{n \in \omega} f(x_{n})  \]
which says that the information produced at the limit of an infinite process of information increase is no more than what can be obtained as the limit of the information produced at  the finite stages of the process.
Note that the ``other half'' of continuity
\[ \bigsqcup_{n \in \omega} f(x_{n}) \sqsubseteq f (\bigsqcup_{n \in \omega} x_{n}) \]
follows from monotonicity.

\subsection{The Fixpoint Theorem}
\begin{theorem}[The Fixpoint Theorem \cite{lassez1982fixed}]
Let $D$ be an $\omega$-cpo with a least element, and $f : D \rightarrow D$ a continuous function. Then $f$ has a least fixed point $\lfp{f}$. Moreover, $\lfp{f}$ is defined explicitly by:
\begin{equation}
\label{lfpeq}
 \lfp{f} = \olub f^{n} (\bot ) . 
\end{equation}

\end{theorem}

This is a central pillar of domain theory in its use in the semantics of computation, providing the basis for interpreting recursive definitions of all kinds \cite{winskel1993formal,gunter1992semantics}.
Note however that the key fixpoint result for computability theory is the \emph{Kleene second recursion theorem} \cite{kleene1938notation,moschovakis2010kleene}, an \emph{intensional} result, which refers to fixpoints for \emph{programs} rather than the functions which they compute.
This theorem is strangely absent from Computer Science, although it can be viewed as strictly stronger than the extensional theorem above.
This reflects the fact which we have already alluded to, that while Computer Science embraces wider notions of processes than computability theory, it has tended to refrain from studying intensional computation, despite its apparent expressive potential. This reluctance is probably linked to the fact that it has proved difficult enough to achieve software reliability even while remaining within the confines of the extensional paradigm. Nevertheless, it seems reasonable to suppose that understanding and harnessing intensional methods offers a challenge for computer science which it must eventually address.

\subsection{Trouble in Paradise}

The semantic theory of higher-type functional computation we get from domain theory seems compelling.
But there is a problem: a mismatch between what the semantic theory allows, and what we can actually compute in `natural' programming languages.

\paragraph{Example: \emph{parallel or}}
We consider a (curried) function of two arguments
\[ \mathsf{por}: \Bool_{\bot} \rarr \Bool_{\bot} \rarr \Bool_{\bot} \]
\[ \por \, \bot \, \true = \true = \por \, \true \, \bot, \qquad \por \, \false \, \false = \false \]
This is the `strong or' of Kleene 3-valued logic.

This function lives in the semantic model, since it is monotonic and continuous, but it is not definable in realistic languages --- or those arising from logical calculi (essentially, the $\lambda$-calculus).
This is because a definable function of two arguments must examine its arguments in some definite order; whichever it examines first, the function will yield an undefined result if that argument is undefined. 
We can define second order functions which can only be distinguished by their values at parallel or; they will be \emph{observationally equivalent}, \ie equivalent in the operational sense, since no experiment we can perform within the language by applying them to inputs which can be defined in the language will serve to distinguish them. However, these second-order functions will have \emph{different denotations} in a model which includes parallel or. Thus such a model --- and in particular, the canonical domain-theoretic model --- introduces operationally unjustified distinctions.

Note that \emph{definability} and \emph{higher types} are crucially important here.
This mismatch between model and operational content is known as the \emph{failure of full abstraction} \cite{milner1977fully,plotkin1977lcf}.

\subsection{Sequentiality}

Consider the following functions, which \emph{are} definable:
\[ \lsor(b_1, b_2) \; \equiv \; \Cond{b_1}{\true}{b_2} \]
\[ \rsor(b_1, b_2)  \; \equiv \; \Cond{b_2}{\true}{b_1} \]
Note that
\[ \lsor(\true, \bot) = \true, \qquad \lsor(\bot, \true) = \bot \]
\[ \rsor(\true, \bot) = \bot, \qquad \rsor(\bot, \true) = \true \]
To get $\por$, we need to run the \emph{processes} for evaluating $b_1$ and $b_2$ in parallel.
For example, we could interleave the executions of these processes, running each one step at a time. As soon as one of these returned the result $\true$, we could return $\true$ as the value of $\por$.
This means that we need to have access, not just to the purely extensional information about the arguments --- \ie their values --- but to their \emph{intensional descriptions}.
This form of \emph{dovetailing} is fundamental to many constructions in computability theory \cite{rogers1987theory}, but is not available in logical calculi and functional languages based on the 
$\lambda$-calculus. Indeed, contemporary functional programming languages take an extensional view of data as one of their cardinal virtues \cite{hughes1989functional,wadler1992essence,turner1995elementary}.

This raises the question: how can we capture the notion of \emph{sequential functional}?

\subsection{Sequentiality is extrinsic}
We need additional information to characterize those functionals which are sequential.
The following remarkable result, due to Ralph Loader \cite{loader2001finitary}, shows that this is unavoidable.

\begin{theorem}[Loader]
The set of functionals definable in Finitary PCF, the typed $\lambda$-calculus over the booleans with conditionals and a term denoting $\bot$, is not recursive.
\end{theorem}

Note that, since the base type here is just the flat domain of booleans, the set of functionals at any type is \emph{finite}. Thus if we form the logical type theory over this structure  by closing under cartesian product and powerset, then all the logical types over this structure will also be finite. If there were any form of intrinsic definition of sequentiality, given in terms of some structure defined at each type even in full higher-order logic, this finiteness would ensure that the notion of sequentiality was recursive. By Loader's theorem, we conclude that no such intrinsic definition can exist.\footnote{This argument is due to Gordon Plotkin (personal communication).}
This rules out any hope of a reasonable intrinsic definition of the sequential functionals.
So the best we can do is to characterize which are the \emph{sequential functional processes} --- \ie an intensional notion.

Attempts at such characterizations coming from a computability theory perspective were made by  Kleene in his series of papers 
%\textsl{Recursive Functionals and Quantifiers of Finite Type Revisited I--IV}  
\cite{kleene1978recursive,kleene1980recursive,kleene1982recursive,kleene1985unimonotone,kleene91},
and by Gandy and Pani in unpublished work.
The work of Berry and Curien on sequential algorithms on concrete data structures \cite{berry1982sequential}, directly inspired by the full abstraction problem for PCF, should also be mentioned.

A characterization of the sequential functional processes was eventually achieved using the newly available tools of
Game Semantics, developed in the 1990's by the present author, Radha Jagadeesan and Pasquale Malacaria, and by Martin Hyland and Luke Ong \cite{abramsky2000full,hyland2000full}. We shall now give a brief account of these ideas.

\section{Game Semantics}

We shall give a brief, informal introduction to game semantics, emphasizing concepts and intuitions rather than technical details, for which we refer to works such as \cite{abramsky2000full,hyland2000full,abramsky1997semantics,abramsky1999game}.

The traditional approach to denotational semantics, exemplified by domain theoretic semantics \cite{gunter1992semantics,winskel1993formal}, was to interpret the types of a programming language by (possibly structured) sets, and the programs as functions.
Game semantics fundamentally revises this ontology:
\begin{itemize}
\item Types of a programming language are interpreted as 2-person games: the {\color{red} Player} is the System (program fragment) currently under consideration, while the {\color{blue} Opponent} is the Environment or context.
\item Programs are \emph{strategies} for these games.
\end{itemize}

So game semantics is inherently a semantics of \emph{open systems}; the meaning of a program is given by its potential interactions with its environment.

A key feature of game semantics as developed in computer science (and significantly differentiating it from previous work on games in logic \cite{lorenzen1960logik,hintikka2010}) is its \emph{compositionality}. The key operation is plugging two strategies together, so that each \emph{actualizes} part of the environment of the other.
The familiar game-theoretic idea of playing one strategy off against another is a special case of this, corresponding to a \emph{closed} system, with no residual environment.
This form of interaction exploits the game-theoretic P/O duality.

\subsection{Types as Games}

\begin{itemize}
\item A simple example of a basic datatype of natural numbers:
\[ \Nat = \{ {\color{blue} q}\cdot{\color{red} n} \mid n \in \mathbb{N} \} \]
Note a further classification of moves, orthgonal to the P/O duality;the O-move  ${\color{blue} q}$ is a \emph{question}, the P-moves ${\color{red} n}$ are \emph{answers}. This turns out to be important for capturing \emph{control features} of programming languages.

\item Forming function or procedure types $A \Rightarrow B$.
We form a new game from disjoint copies of $A$ and $B$, \emph{with P/O roles in $A$ reversed}.
Thus we think of $A \Rightarrow B$ as a \emph{structured interface} to  the Environment; in $B$, we interact with the caller of the procedure, \emph{covariantly}, while in $A$, we interact with the argument supplied to the procedure call, \emph{contravariantly}.

\end{itemize}

\subsection{Example}

We consider the strategy corresponding to the term $\lambda f : \Nat \Rightarrow \Nat. \, \lambda x: \Nat . \, f(x) + 2$.
This term defines the procedure $P(f, x)$ such that $P(f, x)$ returns $f(x) + 2$.

We show a typical play for the strategy corresponding to this term in Figure~1.

\begin{figure}
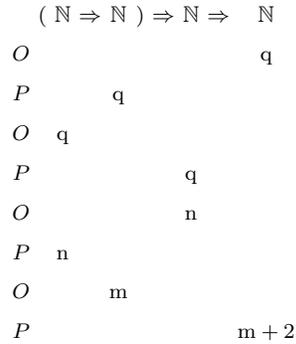

\label{stratfig}
\begin{center}
$\begin{array}{cccccccccc}
& ( & \Nat & \Rightarrow & \Nat & ) & \Rightarrow & \Nat & \Rightarrow & \Nat\\
\Opp&&&&&&&&&{\color{blue} \mathrm{q}}\\
\Pla &&&&{\color{red} \mathrm{q}}\\
\Opp&&{\color{blue} \mathrm{q}}\\
\Pla &&&&&&&{\color{red} \mathrm{q}}\\
\Opp&&&&&&&{\color{blue} \mathrm{n}}\\
\Pla &&{\color{red} \mathrm{n}}\\
\Opp&&&&{\color{blue} \mathrm{m}}\\
\Pla &&&&&&&&&{\color{red} \mathrm{m+2}}\\
\\
\end{array}$
\end{center}
\caption{Strategy for $\lambda f : \Nat \Rightarrow \Nat. \, \lambda x: \Nat . \, f(x) + 2$.}
\end{figure}
This should be read as follows. Time flows downwards. Moves by Opponent  alternate with those of Player. Each column corresponds to one of the occurrences of atomic types in the overall type of the term. Each of these will be a copy of the simple game for natural numbers. In effect, we are playing on several game boards, switching from one to another.

The play begins with the Opponent, or environment, requesting an output.
The strategy must call the function argument $f$, so it requests an output from this argument. Note that the variance rules dictate that, since the type of $f$ occurs negatively in the overall type, this opening move is indeed a Player move.
The environment now has to respond to this request. Typically, it will do so by requesting its input. Since the strategy is realizing the function call $f(x)$, the value of this input will be obtained by the strategy from the natural number argument $x$, and thus the strategy requests the value of this argument. When the evironment responds to this request with some number $n$, the strategy \emph{copies} this value to answer the request of the function argument for its input. This copying of data is a key feature of game semantics, corresponding to the logical flow of information. Indeed, purely logical strategies, \ie those corresponding to logical proofs, are typically made entirely from such `copy-cat' processes \cite{abramsky1994games}.

After the function argument receives its input $n$, it will typically emit some output $m$; the strategy for the term will now answer the original request from the environment with the number $m+2$.

We have simply described what the evident strategy corresponding to the above $\lambda$-term will do.
Of course, the purpose of a compositional game semantics is precisely to allow this strategy to be constructed systematically as the denotation of the above term in a syntax-directed fashion.

\subsection{Composition}

We now show how the key operation of plugging one strategy together with another, corresponding to applying a procedure to its argument, or more generally to composing procedures, will proceed, again through an example.
We apply the higher-order procedure $\lambda f : \Nat \Rightarrow \Nat. \, \lambda x: \Nat . \, f(x) + 2$
to the argument $\lambda x:\Nat . \, x^2$.
A typical run of the corresponding strategy is shown in Figure~2.

\begin{figure}
\begin{center}
$\begin{array}{cccccccccccccc}
\rnode{tl}{} & & & & \rnode{tr}{}  & \rnode{ul}{} & & && \rnode{ur}{} &&&&  \\
&\Nat & \Rightarrow & \Nat & \;\; & ( & \Nat & \Rightarrow & \Nat & ) & \times & \Nat & \Rightarrow & \Nat\\
&&&&&&&&&&&&&{\color{blue} \mathrm{q}}\\
&&& {\color{blue} \mathrm{q}} &&&&&{\color{red} \mathrm{q}}\\
& {\color{red} \mathrm{q}} &&&&&{\color{blue} \mathrm{q}}\\
&&&&&&&&&&&{\color{red} \mathrm{q}}\\
&&&&&&&&&&&{\color{blue} \mathrm{n}}\\
& {\color{blue} \mathrm{n}}&&&&&{\color{red} \mathrm{n}}\\
&&&{\color{red} \mathrm{n^2}}&&&&&{\color{blue} \mathrm{n^2}}\\
&&&&&&&&&&&&&{\color{red} \mathrm{n^2} + 2} \\
\rnode{bl}{} & & & & \rnode{br}{} & \rnode{cl}{} & & & & \rnode{cr}{} &&&&
\ncline[linestyle=dotted]{tl}{tr}
\ncline[linestyle=dotted]{bl}{br}
\ncline[linestyle=dotted]{tl}{bl}
\ncline[linestyle=dotted]{br}{tr}
\ncline[linestyle=dotted]{ul}{ur}
\ncline[linestyle=dotted]{cl}{cr}
\ncline[linestyle=dotted]{ul}{cl}
\ncline[linestyle=dotted]{cr}{ur}
\end{array}$
\end{center}
\caption{Applying $\lambda f : \Nat \Rightarrow \Nat. \, \lambda x: \Nat . \, f(x) + 2$
to $\lambda x:\Nat . \, x^2$}
\end{figure}

Here the common parts of the two strategies are shown in the dashed boxes. The key point is that type matching between function and argument guarantees that the Opponent moves for one are Player moves for the other, and vice versa. This intrinsic duality allows the interactive interpretation of composition to be defined without requiring any additional structure. Thus when the strategy for the functional plays its opening  move in response to the environment request for an output, the other strategy sees this as the opening  move of \emph{its} environment, and responds accordingly. The functional now sees this response as the next move of the nvironment, and so on. With each of the strategies following their own scripts, we get a uniquely determined path through the common part of the type, where they interact.

Note that the \emph{residual strategy}, after we hide the part where the interaction between function and argument occurs, is that corresponding to the function $\lambda x.\, x^2 + 2$, consistent with the result of performing $\beta$-reduction syntactically. This `parallel composition plus hiding' paradigm for composition of strategies \cite{abramsky1994games} is a fundamental component of game semantics.

\subsection{Technical Notes}
Once the ideas shown informally by example in the previous subsection are properly formalized, 
games and strategies organize themselves into a very nice mathematical structure --- a cartesian closed category $\GG$ \cite{abramsky2000full,hyland2000full}. We can then use standard methods of denotational semantics to give a compositional semantics for functional languages such as PCF in $\GG$\cite{gunter1992semantics}.

The key result is:

\begin{theorem}[Abramsky-Jagadeesan-Malacaria and Hyland-Ong \cite{abramsky2000full,hyland2000full}]
\label{AJMHOthm}
Every compact (even: recursive) strategy in $\GG$ is definable by a PCF term.
\end{theorem}

This can be understood as a \emph{completeness theorem}. It is saying, not only that every sequential functional process can be interpreted as a strategy in the game semantics, but that the `space' of strategies corresponds \emph{exactly} to the class of sequential functional processes.
In short, we have achieved a characterization, albeit at the intensional level.

Note that similar results can be achieved for proof calculi for various logical systems, leading to notions of \emph{full completeness} \cite{abramsky1994games,abramsky1999concurrent}.

If we refer to strategies as in our examples in the previous sub-section, we can get some intuition for how a result such as this is proved. The
first move (by Player) in the strategy corresponds to the  head variable in the head-normal form of the $\lambda$-term, or the last rule in a proof. Decomposing the strategy progressively uncovers the defining term. This argument is formalized in \cite{abramsky2000full,hyland2000full}, and even axiomatized in \cite{abramsky1999axioms}.

There is an important caveat, which actually leads to a major positive feature of game semantics.
In order to achieve this  completeness result, strategies must be  \emph{constrained}, as we shall now explain. 
%This is our first major example of sub- (or non-) expressiveness.

\subsection{Constraints on Strategies}

There are two main kinds of constraints which must be imposed on strategies in order to capture exactly the sequential functional processes:

\begin{enumerate}
\item Strategies do not have perfect information about the previous computation history (play of the game).
For example, consider a call
\[ f t_1 t_2 \]
When it is evaluated (called) by $f$,  $t_2$ should not `know' whether $t_1$ has already been evaluated --- and vice versa.
Note that if we had imperative state, the arguments could use this to pass such information to each other. Thus this constraint is a distinctive feature of purely functional computation.

\item Properly nested call-return flow of control. This is visibly a feature of the example we discussed previously. It is often referred to as the `stack discipline' \cite{abramsky2000full}, for obvious reasons. It corresponds to the absence of  non-local control features such as jumps or exceptions. Again, this constraint is a distinctive feature of purely functional computation.
\end{enumerate}

Both these constraints can be formulated precisely (the first as `innocence' or `history-freedom', the second as `well-bracketing'), and shown to be closed under composition and the other semantic constructions \cite{abramsky2000full,hyland2000full}. 
The resulting cartesian closed categories of games and constrained strategies yield the completeness results as in Theorem~\ref{AJMHOthm}.

\subsection{Discussion}

It should be noted that there is an important caveat to the result in Theorem~\ref{AJMHOthm}.
 We get a definability result --- but strategies are much finer grained than functions. They correspond to certain `PCF evaluation trees' \cite{abramsky2000full}. To get an equationally fully abstract model, we must quotient the games model.
This is unavoidable by Loader's Theorem, from which it follows that the fully abstract model, even for Finitary PCF, is \emph{not effectively presentable}.

What is the significance of the result? In retrospect, the real payoff is  in other cases --- but the PCF result is the keystone of the whole development.
Relaxing the constraints which characterize functional computation leads to fully abstract models for languages with (locally scoped) state, or control operators, or both \cite{abramsky1999game}. This picture of a semantic `cube' or hierarchy of constraints has proved very fruitful as a paradigm for exploring a wide range of programming language features.

\subsection{The Game Semantics Landscape}

Game semantics has proved to be a flexible and powerful paradigm for constructing highly structured fully abstract semantics for languages with a wide range of computational features:
\begin{itemize}
\item (higher-order) functions and procedures \cite{abramsky2000full,hyland2000full}
\item call by name and call by value \cite{honda1997game,abramsky1998call}
\item locally scoped state \cite{abramsky1997linearity,murawski2009full}
\item general reference types \cite{abramsky1998fully,murawski2011game}
\item control features (continuations, exceptions) \cite{laird2003game,laird2001fully}
\item non-determinism, probabilities \cite{harmer1999fully,danos2002probabilistic}
\item concurrency \cite{laird2005game,laird2001game,ghica2008angelic}
\item names and freshness \cite{abramsky2004nominal,tzevelekos2007full}.
\end{itemize}

In many cases, game semantics have yielded the first, and often still the only, semantic construction of a fully abstract model for the language in question.
Moreover, where sufficient computational features (typically state or control) are present, then the observational equivalence is more discriminating, and  game semantics captures the fully abstract model directly, without the need for any quotient.
Intensions become extensions!

More generally: the point of conceptual interest is to find \emph{positive reasons} --- structural invariants --- \emph{for non-expressiveness}. 
This is a positive story for a form of intensionality.
Indeed, we can see here the beginnings of a \emph{structural theory of processes}.

\subsection{Mathematical Aside}
Categories of games and strategies have fascinating mathematical structure in their own right.
They give rise to:
\begin{itemize}
\item Constructions of \emph{free categories with structure} of various kinds.
\item \emph{Full completeness} results characterizing the ``space of proofs'' for various logical systems \cite{abramsky1994games,abramsky1999concurrent}.
\item There are even connections with \emph{geometric topology}, e.g.~Temperley-Lieb and other diagram algebras \cite{abramsky2009temperley}.
\end{itemize}

\subsection{Algorithmic Game Semantics}

We can also take advantage of the \emph{concrete nature} of game semantics.
A play is a sequence of moves, so a strategy can be represented by the set of its plays, i.e. by a \emph{language} over the alphabet of moves, and hence by an automaton.

There are significant finite-state fragments of the semantics for various interesting languages, as first observed by Ghica and McCusker \cite{ghica2003regular,abramsky2002algorithmic}.
This means we can \emph{compositionally construct} automata as (representations of) the meanings of open (incomplete) programs, giving a powerful basis for compositional software model-checking.
There has been an extensive development of these ideas in the last few years, by Ghica, Ong, Murawski, Tzevelekos \textit{et al.}~\cite{abramsky2003algorithmic,legay2008automated,ghica2009applications,tzevelekos2012alg}.

\subsection{Other Aspects}

It should also be noted that there are other strands in  game semantics, which has become a rich and diverse field.
For example, there is work on clarifying the mathematical structure of game semantics itself, and its relation to syntax and to categorical structure. Examples include \cite{mellies2006asynchronous,mellies2012game,harmer2007categorical,winskel2012bicategories,curien2012approach}.

More broadly, we must emphasise that our subject in this paper has been the characterisation of sequential, functional processes. We have argued that a considerable measure of success has been achieved in answering this question. However, finding a compelling answer to the general question of ``What is a process?'' which we raised in the Introduction remains a major challenge of the field. It is likely that new conceptual ingredients will be needed to enable further advances towards this goal.

\section{Conclusions: Some Questions and Dreams}

\begin{itemize}
\item Can we use intensional recursion to give more realistic models of reflexive phenomena in biology, cognition, economics, etc.?

Working with codes seems more like what biological or social mechanisms might plausibly do, rather than with abstract mathematical objects \textit{in extenso}.

\item How does this relate to current interest in higher categories, homotopy type theory etc. \cite{awodey2007homotopy,voevodsky2010univalent}, where equalities are replaced by \emph{witnesses}? Do higher categories, and the intensional type theories they naturally support, provide the right setting for a systematic intensional view of mathematics and logic? What kind of novel applications will these structures support?

\item Can we develop a positive theory of intensional structures, and find the right notions of invariance?

\item The dream: to use this to give some (best-possible) \emph{intrinsic} characterization of computability, and of complexity classes.

\item Could this even be the missing ingredient to help us separate classes? Well, we can dream!
\end{itemize}

\bibliographystyle{alpha}

\bibliography{bibfile}

\end{document}